%% file: letter.tex
\documentclass[conference]{IEEEtran}
\IEEEoverridecommandlockouts

\usepackage[english]{babel}
\usepackage[T1]{fontenc}
\usepackage{amssymb, amsmath, amsthm}
\usepackage{mathtools}
\usepackage{cite}
\usepackage{graphicx}
\usepackage{subfloat}
\usepackage[caption=false,font=footnotesize,farskip=0pt]{subfig}
\usepackage{xcolor}
\usepackage{multirow}
\usepackage[ruled,vlined,linesnumbered]{algorithm2e}
\usepackage{enumitem}
\usepackage{stfloats}
\usepackage[normalem]{ulem}
\usepackage{stix}
\usepackage{booktabs}
\usepackage{tabularx}
\usepackage{multirow}
\usepackage{multicol}
\usepackage{amsmath}
\usepackage{algorithmic}

\usepackage{xcolor, colortbl}
\usepackage[acronym,,nohypertypes={acronym,notation}]{glossaries}
\input{acronym}

\definecolor{cian}{rgb}{.02,.7,.95}
\definecolor{gold}{rgb}{0.85,.66,0}

\DeclareMathOperator*{\argmax}{arg\,\max}

\title{
Proximal Policy Optimization for Integrated Sensing and Communication in mmWave Systems
\thanks{
$^*$Corresponding Author. This work has been submitted to IEEE for possible publication. Copyright may be transferred without notice, after which this version may no
longer be accessible.
}
}

\author{
    \IEEEauthorblockN{
        {Cristian J. Vaca-Rubio$^*$},
        {Carles Navarro Manchón},
        {Ramoni Adeogun},
        and {Petar Popovski} \\
        }
        \IEEEauthorblockA{
        \textit{Department of Electronic Systems, Aalborg University}, Aalborg, Denmark\\
        E-mail:\{cjvr, cnm, ra, petarp\}@es.aau.dk
        }
}

\begin{document}

\maketitle

\begin{abstract}
In wireless communication systems, mmWave beam tracking is a critical task that affects both sensing and communications, as it is related to the knowledge of the wireless channel. We consider a setup in which a Base Station (BS) needs to dynamically choose whether the resource will be allocated for one of the three operations: sensing (beam tracking), downlink, or uplink transmission.
We devise an approach based on the Proximal Policy Optimization (PPO) algorithm for choosing the resource allocation and beam tracking at a given time slot.
The proposed framework takes into account the variable quality of the wireless channel and optimizes the decisions in a coordinated manner. Simulation results demonstrate that the proposed method achieves significant performance improvements in terms of average packet error rate (PER) compared to the baseline methods while providing a significant reduction in beam tracking overhead. We also show that our proposed PPO-based framework provides an effective solution to the resource allocation problem in beam tracking and communication, exhibiting a great generalization performance regardless of the stochastic behavior of the system.
\end{abstract}

\begin{IEEEkeywords}
mmWave, sensing, beam tracking, resource allocation.
\end{IEEEkeywords}
\vspace{-2mm}
\section{Introduction}
Beamforming and resource allocation methods have significantly improved as a result of the rising demand for fast and dependable wireless communication networks. A crucial task in wireless communication systems is beam alignment, which entails directing antenna arrays to maximize throughput and improve signal-to-noise ratio (SNR). Traditional beam sweeping in 5G  allocates sensing resources for all users at fixed-slots periodicity. This might lead to inefficient (low-mobility scenarios) and insufficient (high-mobility) to track the necessity of beam changes. These operations are not typically adapted at short time-scales\cite{giordani2018tutorial}, and the division between uplink (UL) and downlink (DL) slots is usually kept fixed over long time frames. To address this, we aim to find a way to decide the allocation of resources between sensing, uplink, and downlink transmissions that is adaptive on a short time-scale according to the particular network situation.


In recent years, beam alignment issues in wireless networks have been successfully resolved using reinforcement learning (RL) techniques \cite{susarla2022hierarchial, raj2022deep}, as well as pure deep learning (DL) solutions \cite{ma2020machine, wang2019mmwave}. The latter rely on the one-shot interaction with the channel, making it more prone to generalization issues, while the former do not consider the joint problem of resource allocation and beam alignment. Similarly, resource allocation is a critical task in wireless communication systems, and it has been traditionally formulated as an optimization problem~\cite{liang2019deep}. However, the increasing complexity of wireless networks requires techniques beyond traditional optimization, such as Reinforcement learning (RL). For instance, RL-based methods have been proposed for joint power and radio resource allocation in 5G networks \cite{liang2019deep, sridhara2008spectrum}. 


Motivated by the encouraging findings of previous works, we propose a Proximal Policy Optimization (PPO) \cite{schulman2017proximal} solution for joint resource allocation and beam tracking in a mmWave wireless communication context. By coordinating the optimization of the resource allocation and beam tracking parameters, our system intends to mitigate the wireless channel's dynamic nature and its fluctuating channel quality. Simulation results show that, when compared to baseline approaches, the suggested method significantly outperforms them in terms of average packet error rate (PER), learning a more dynamic policy of slots allocation for efficient beam tracking and data transmission.
\vspace{-2mm}
\section{System Model and Problem Formulation}
\begin{figure}[t]
    \centering
    \includegraphics[width=0.7\columnwidth]{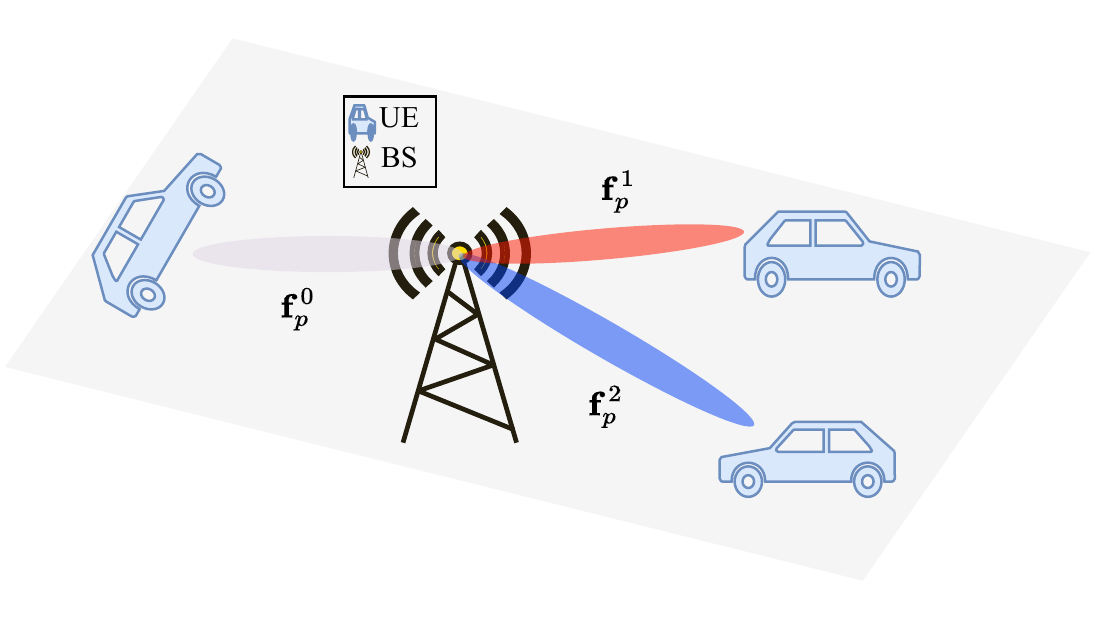}
    \caption{Illustration of the scenario}
    \label{fig:system}
\end{figure}

Consider a vehicular wireless network as depicted in Figure \ref{fig:system}, using a time division duplex (TDD) scheme. A mmWaveBase Station (BS) equipped with a uniform linear array (ULA) with $M_t$ transmitting antennas communicates with $U$ single-antenna User Equipment (UEs) in a time-slotted network with $\mathcal{K} = \{1,...,K\}$ orthogonal channel uses per frame. In each time slot, only one user $u$ can be scheduled. Each user $u$ has a buffer of size $B^u_{UL}$ for uplink packets, while the base station has U buffers (with size $B^u_{DL}$ for the user u) for downlink traffic. At each slot $k \in \mathcal{K}$, the BS performs one action: $S_u$ (sensing/beam tracking), $DL_u$ (transmission of downlink packet), or $UL_u$ (reception of uplink packet) with $u = {1, ..., U}$.
\subsection{Channel Model}
Due to the high directivity of mmWave systems, we assume a geometrical Line-of-Sight (LoS) channel model for this work. In this way, the narrowband channel for the user $u$ is given by
\begin{equation}
    \mathbf{h}_u = \frac{\sqrt{M_t}}{d_u}\beta\mathbf{a}_t^\dagger(\theta_u),
\end{equation}
where $\dagger$ denotes conjugate transposition, $\mathbf{h}_u \in \mathbb{C}^{M_t}$ denotes the channel vector for user $u$, $d_u$ denotes the distance between the BS and the $u-$th user and $\beta \sim \mathcal{CN}(0, \sigma_u^2)$ is the complex fading gain with variance $\sigma_u^2$. Also, $\theta_u$ denotes the Angle of Departure (AoD) with respect to the BS array axis and the $u$ user position. As we are considering a ULA BS, the ideal isotropic array responses are given by
\begin{equation}
    \mathbf{a}_t(\theta_u) = \frac{1}{\sqrt{M_t}}\left[1, e^{-j\pi\cos(\theta_u)}, ..., e^{-j\pi(M_t-1)\cos(\theta_u)}\right]^T.
\end{equation}
For the dynamics of the system \cite{raj2022deep}, given the user position in a given time slot $\mathbf{c}^u_k=[x^u_k, y^u_k]$, its velocity $\mathbf{v}=[v_\text{linear}, v_\text{angular}]$, where $v_\text{linear} \sim \text{Exp}(1)$ and $v_\text{angular} \sim \mathcal{N}(0, 1)$ and a slot duration of $\Delta t$, the user new position can be calculated as follows: \textit{i)} compute the angular displacement,  $\omega = v_\text{angular} \Delta t$, \textit{ii)} compute the direction vector: $\mathbf{d} = [\cos \omega, \sin \omega]$, \textit{iii)} compute the linear displacement: $\mathbf{\Delta c} = v_\text{linear} \mathbf{d} \Delta t$, and \textit{iv)} compute the user's new position: $\mathbf{c}^u_{k+1} = \mathbf{c}^u_k + \mathbf{\Delta c}$. This will change the AoDs and consequently make the channel vary along our time-slotted resources. For simplicity, we assume a 2D geometric layout.
\subsection{Beam Codebook}
In this work, we assume a codebook-based analog beamforming architecture to beamform signals with a single RF chain at the BS. We denote by $\mathcal{F} = \{\mathbf{f}_1, ..., \mathbf{f}_{M_t}\}$ the codebook used for analog beamforming at the BS, with $M_t$ beams. We use the common Discrete Fourier Transform (DFT)-based codebooks \cite{rezaie2020location}, with precoders $\mathbf{f}_i$ given by
\begin{equation}
    \mathbf{f}_i = \frac{1}{\sqrt{M_t}}\left[1, e^{-j\pi\frac{2i-1-M_t}{M_t}}, ..., e^{-j\pi(M_t-1)\frac{2i-1-M_t}{M_t}}\right]^T,
\end{equation}
where $i \in \{1, ..., M_t\}$.
Then on each time slot, given the selected precoder $\mathbf{f}^u_i \in \mathcal{F}$, the received signal power $R^u_i \in \mathbb{R}$ for the $u$-th user can be described as
\begin{equation}
    R^u_i = |\sqrt{P_t}\mathbf{h}^T_u\mathbf{f}^u_is+ \mathbf{n}_u|^2,
\end{equation}
where $T$ denotes transposition, $P_t$, $s \in \mathbb{C}$, and $\mathbf{n}_u$ denote the transmission power, the known training symbol with normalized power, and the zero mean complex Gaussian noise vector with variance $\sigma_n^2$.
\subsection{Traffic model}
Each user follows an independent Bernoulli process with probability $P^u$ for generating packets in downlink and uplink in every time slot $k$.
\begin{equation}
    [p^u_{dl}, p^u_{ul}] \sim \text{Bernoulli}(P^u),
\end{equation}
where $p^u_{dl}, p^u_{ul}$ denote 0 or 1 to determine if a packet was generated for the user $u$ in downlink and uplink for a time slot $k$.
\subsection{Optimization Problem}
We perform the allocation of slots for a specific user $u$ with the goal of minimizing the average PER, defined as:
\begin{equation}
    PER = \frac{Lost Packets}{Total Packets}
\end{equation} 
within an episode $\mathcal{K}$. The optimization problem is formulated as follows:
\begin{align}
\min_{k \in \mathcal{K}} \quad & \frac{1}{UK} \sum_{u=1}^{U} \sum_{k=1}^{K}  PER_{uk}\\
\text{s. t.} \quad & \sum_{u=1}^{U} s_{uk} + \sum_{u=1}^{U} d_{uk} + \sum_{u=1}^{U} u_{uk} = 1, \quad \forall k=1,2,...,K \notag \\
& s_{uk}, d_{uk}, u_{uk} \in \{0,1\}, \quad \forall u=1,2,...,U,\ k=1,2,...,K \notag
\end{align}
where the constraints denote that just one user can be allocated in every slot $k$ for sensing $s_{uk}$, downlink $d_{uk}$or uplink $u_{uk}$, respectively.
\section{PPO Fundamentals}
PPO is an actor-critic algorithm \cite{schulman2017proximal} that models both the policy and value functions as a neural network, with parameters denoted as $\theta$, and $\phi$. It aims to learn a policy $\pi_{\theta}(a|s)$ that maximizes the expected cumulative reward in an environment. The algorithm utilizes a value function $V_{\phi}(s)$ to estimate the expected total reward from a given state $s$, and an advantage function $A_k$ to measure the quality of an action in a specific state. 
At each time step $k$, the agent observes the current state $s_k$ of the environment and samples an action $a_k$ from the policy distribution $\pi_{\theta}(a_k|s_k)$. The action is then executed, resulting in a reward $r_k$ and a new state $s_{k+1}$. The agent stores these experiences to update its policy and value functions after accumulating $N$ tuples of experience for a given policy.  From here, we will denote the equations for a single time step $k$ for simplicity. The advantage function $A_k$ measures the advantage of taking action $a_k$ in state $s_k$ compared to the expected value from the current state. It is computed as the sum of the discounted future rewards minus the value function at the current state:
\begin{equation}
    A_k = \sum_{i=k}^{N} \gamma^{i-k} r_i - V_{\phi}(s_k),
\end{equation}
where $N$ is the maximum number of time steps per experience memory. Next, the value function $V_{\phi}(s_k)$ estimates the expected cumulative reward from the current state $s_k$ onwards. It is updated by minimizing the mean squared error between the estimated value and the target value:
\begin{equation}
    L_{\text{critic}}^{(k)}(\phi) = \frac{1}{2}\left(V_{\phi}(s_k) - (r_k + \gamma V_{\phi}(s_{k+1}))\right)^2,
\end{equation}
where $\gamma$ is the discount factor that balances immediate and future rewards.

The policy is updated using the PPO objective, which aims to maximize the expected advantage while avoiding large policy changes. The PPO loss function for a single time step is given by:
\begin{equation}
    L_{\text{PPO}}^{(k)}(\theta) = \min\left(r_k(\theta)A_k, \text{clip}(r_k(\theta), 1 - \epsilon, 1 + \epsilon)A_k\right),
\end{equation}
where the $\text{clip}$ operation restrict the values from $1 - \epsilon$ to $1 + \epsilon$,  $r_k(\theta) = \frac{\pi_{\theta}(a_k|s_k)}{\pi_{\text{old}}(a_k|s_k)}$ is the ratio of the updated policy probabilities to the old policy probabilities, and $\epsilon$ is a hyperparameter that controls the magnitude of the policy change.
To further improve the stability of training, an entropy regularization term is included. The entropy $S[\pi_{\theta}(a_k|s_k)]$ measures the uncertainty or randomness in the policy distribution:
\begin{equation}
    S[\pi_{\theta}(a_k|s_k)] = -\sum_{a} \pi_{\theta}(a_k|s_k) \log \pi_{\theta}(a_k|s_k).
\end{equation}
This term encourages exploration by discouraging overly deterministic policies.
Finally, the total loss function for PPO combines the policy loss, value function loss, and entropy regularization over the entire $N$ experiences:
\begin{equation}
L_{\text{total}}(\theta, \phi) = \frac{1}{N}\sum_{k=1}^{N} \left( L_{\text{PPO}}^{(k)}(\theta) - c_1L_{\text{critic}}^{(k)}(\phi) + c_2S[\pi_{\theta}(a_k|s_k)] \right)
\end{equation}
where $c_1$ and $c_2$ are hyperparameters that control the trade-off between the value function loss and the entropy regularization. The policy and value networks are then updated jointly by minimizing the total loss. 

\begin{algorithm}[t]
\KwIn{The policy network $\pi_{\theta}(a, s)$, the value network $V_{\phi}(s)$ with weights $\theta$ and $\phi$, $episodes$, learning rate $\alpha$, discount factor $\gamma$, weighting factor $\lambda$, number of time steps per episode $K$, environment $env$, clipping parameter $\epsilon$}
\KwOut{Policy function $\pi_{\theta}$, value function $V_{\phi}$}
Initialize $\theta$ and $\phi$ weights \\
\For{$i \leftarrow 1 \textbf{ to } episodes$}{
Initialize environment: $s_1 \gets env.reset()$\\
Initialize empty lists for log probabilities, values, rewards, and old policy probabilities \\
Initialize counter $c \leftarrow 0$ \\
\While{$c < K$}{
\For{$k \leftarrow (1+c) \textbf{ to } (N + c)$}{
Compute action probabilities $\pi_\theta(a_k|s_k)$ and value estimate $V_\phi(s_k)$ from neural network \\
Sample action $a_k \sim \pi_\theta(a_k|s_k)$ and compute log probability $\log \pi_\theta(a_k|s_k)$ \\
Compute old policy probability $\pi_{\text{old}}(a_k|s_k)$ \\
Take action $a_k$ and observe reward $r_k$ and next state $s_{k+1}$ \\
Append log probability $\log \pi_\theta(a_k|s_k)$, value estimate $V_\phi(s_k)$, reward $r_k$, and old policy probability $\pi_{\text{old}}(a_k|s_k)$ to their respective lists \\
Update state to $s_{k+1}$
}
\For{$k \leftarrow (1+c) \textbf{ to } (N+c)$}{
Compute advantages $A_k$\\
Compute ratio $r_t(\theta) = \frac{\pi_\theta(a_k|s_k)}{\pi_{\text{old}}(a_k|s_k)}$ \\
Compute surrogate objective $\hat{L}_{\text{PPO}}(\theta) = \min\left(r_t(\theta)A_k, \text{clip}(r_k(\theta), 1 - \epsilon, 1 + \epsilon)A_k\right)$ \\
Compute value function loss $L_{\text{critic}}^{(k)} = \frac{1}{N} \sum_{k=1+c}^{N+c} (G_k - V_\phi(s_k))^2$\\
Compute entropy regularization term $S[\pi_{\theta}(a_k|s_k)]$ \\
Compute total loss $L_{\text{total}}(\theta, \phi) = \frac{1}{N}\sum_{k=1+c}^{N+c} L_{\text{PPO}}^{(k)}(\theta) - c_1L_{\text{critic}}^{(k)}(\phi) + c_2S[\pi_{\theta}(a_k|s_k)]$\\
Update neural network parameters by minimizing $L_{\text{total}}$ using the optimizer \\
}
$c \leftarrow (N + c)$
}}
\KwRet{$\pi_{\theta}$, $V_{\phi}$}
\caption{PPO Training Procedure}
\end{algorithm}

\section{PPO for joint beam alignment and resource allocation}
To solve the problem in equation (7), the PPO framework is used, with the BS as the agent. To do so, the environment, state space, action space, and reward signal are defined. Algorithm~1 provides a summary of the training procedure.
\subsection{State space}
The state space of this environment is characterized by four parameters:
\begin{itemize}
    \item Current beam indexes $i^u \in \mathcal{I}$ for the $u \in U$ users.
    \item Current amount of packets of downlink buffer $P^u_{DL}$ for the $u \in U$ users.
    \item Current amount of packets of uplink buffers $P^u_{UL}$ for the $u \in U$ users.
    \item Current received power for every selected beam for the $u \in U$ users, $R^u_p$.
\end{itemize}
In this way, the state in time slot $k$ can be represented as
\begin{equation}
\begin{split}
\mathcal{S}_k = \{\{i^1, \ldots, i^U\}, \{P^1_{DL}, \ldots, P^U_{DL}\}, \\ \{P^1_{UL}, \ldots, P^U_{UL}\}, \{R^1_p, \ldots, R^U_p\}\},
\end{split}
\end{equation}
We normalize the beam indexes and buffer values dividing by the number of total beams in the codebook and the buffer sizes, respectively.
 \subsection{Action space}
The action space is the set of all possible actions that the BS agent can choose from at each time slot. While the method presented here can be applied to the selection of any wireless resource, we consider the allocation of time slots. In this way, the possible actions are a discrete variable with $U\times 3$ possible values 
\begin{equation}
    a_k \in \mathcal{A} = \{ (a,u), a = 0,1,2, u = 1, 2, ..., U\},
\end{equation}
with the first element $a$ denoting the action (0 for sensing, 1 for UL transmission, 2 for DL transmission) and the second element $u$ denoting the target user.
Once in a sensing slot $s_{uk}$, we assume we select the best neighboring beam (beam tracking) such that
\begin{equation}\label{eq:beam_selection}
    \argmax_{\mathbf{f}_i^u \in \{\mathbf{f}_{i-1}^u, \mathbf{f}_{i}^u, \mathbf{f}_{i+1}^u\}} |\sqrt{P_t}\mathbf{h}^T_u\mathbf{f}^u_is+ \mathbf{n}_u|^2.
\end{equation}
If the sensing variable $s_{uk}$ is active (i.e., $s_{uk} = 1$), the beam selection for either downlink transmission ($d_{uk}$) or uplink transmission ($u_{uk}$) in slots will be based on the beam selected during the last sensing slot. 
\subsection{Reward signal}
The objective stated in equation (7) is to minimize the average PER. In our problem, packets can be dropped due to different reasons: either encountering a full buffer ($P^u_{DL} > B^u_{DL}$ or $P^u_{UL} > B^u_{UL}$), or transmitting with a poor quality beam (i.e., using an outdated beam $\mathbf{f}^u_i$ obtained from equation \eqref{eq:beam_selection} during $s_{uk}$). A beam is considered outdated, if $\mathbf{f}^u_i$ is no longer the optimal beam. To encourage the reduction of these packet drops, we assign a reward in each time slot $k$ based on the selected action. To understand our reward, let's first define $G^u_b(k)$ as an indicator that equals 1 if, at the end of time step $k$, the buffer size $P^U_b$ for user $u$ (where $b$ can be DL or UL) is such that $P^U_b < B^u_{b}$. This indicates that the buffer is not full. Additionally, we use the factor $\rho(k)$ to represent the impact of beam tracking on the beam. When beam tracking is performed and it results in a change to the beam, $\rho(k)$ is greater than zero ($\rho(k) > 0$). Conversely, when beam tracking is unnecessary and the beam remains unchanged, $\rho(k)$ is less than zero ($\rho(k) < 0$). The purpose of $\rho(k)$ is to encourage the use of sensing to recover the best beam when it is beneficial, and to penalize unnecessary sensing. If it is not a sensing slot, $\rho(k)$ is set to zero ($\rho(k) = 0$). Finally, we denote $D(k)$ as the amount of packets dropped in a time slot $k$. Furthermore,  the reward $r_k$ is defined as
\begin{equation}
    r_k = \frac{1}{U\times 2} \sum_{u=1}^U\sum_{b\in \{DL, UL\}} G^u_b(k) + \rho(k) - D(k),
\end{equation}
\section{Simulation environment and evaluation description}
The algorithm is trained for a fixed number of episodes $E_{\text{train}}$, with some randomly selected initial positions for the UEs with varying speeds, small variations, and arrived packet probabilities for each user. The results are evaluated on $E_{\text{test}}$ randomly initialized episodes to demonstrate generalization capability. Simulation and PPO parameters are listed in Tables I and II, respectively. The baselines used to evaluate the results are described below:
\begin{enumerate}
\item Random: The policy of the agent is to allocate slots uniformly at random among all possible actions.
\item X-TDMA (X Time Division Multiple Access) is a slot allocation policy where slots are divided into sensing (S), uplink (UL), and downlink (DL) categories. The value of X represents the number of consecutive UL/DL slots before transitioning to the next sensing slot. Specifically:
    \begin{enumerate}
    \item For X = 1: The pattern is S-UL-DL, repeating cyclically for each user.
    \item For X = 3: The pattern is S-UL-DL-UL-DL-UL-DL, repeating cyclically for each user
    \item For X = 6: The pattern is S-UL-DL-UL-DL-UL-DL-UL-DL-UL-DL-UL-DL, repeating cyclically for each user.
    \end{enumerate}
In each case, X-TDMA follows TDMA principles and allows users to take turns using the channel resources for uplink and downlink communication. These patterns are allocated for every user $u$.
\end{enumerate}
\vspace{-3mm}
\begin{table}[htbp]
  \centering
  \caption{Simulation Parameters}
    \begin{tabular}{ccc}
    \toprule
    \textbf{Parameter} & \textbf{Symbol} & \textbf{Value} \\
    \midrule
    BS antennas & $M_t$ & 32 \\
    Frequency & $f_c$ & 28 GHz \\
    Tx Power & $P_t$ & 5 W \\
    Signal-to-Noise Ratio & SNR & 20 dB \\
    Scenario dimensions & SD & [100 x 100]  \\
    Number of UEs & U & 3 \\
    Packet arrival probability & $P^u$ & [0.6, 0.4, 0.3] \\
    Packet dl/ul probability & $[p^u_{DL}, p^u_{UL}]$& [0.5, 0.5]\\
    UEs initial positions [(x, y) permute] & $[[x^u_0, y^u_0]]$ & [[0, 80], [0, 40], [0, 27]] \\
    Size of downlink buffers & $B^u_{DL}$ & 5 \\
    Size of uplink buffers & $B^u_{UL}$ & 5 \\
    Number of time slots & $K$ & 1000 \\
    Number of training episodes & $E_{\text{train}}$ & 3000000 \\
    Number of testing episodes & $E_{\text{test}}$ & 6000 \\
    \bottomrule
    \end{tabular}%
  \label{tab:sim_params}%
\end{table}%
\vspace{-6mm}
\begin{table}[htbp]
  \centering
  \caption{Training Algorithm Parameters}
    \begin{tabular}{cc}
    \toprule
    \textbf{Parameter} & \textbf{Value} \\
    \midrule
    Number of layers for actor and critic & 3 \\
    Neurons of actor layers & $[\{state\_size, 64\}, \{64, 64\}, \{64, action\_size\}]$ \\
    Neurons of critic layers & $[\{state\_size, 64\}, \{64, 64\}, \{64, 1\}]$ \\
    Activation function all layers & tanh \\
    Actor probability mapping & softmax \\
    Memory size $N$ & 80 \\
    Learning rate actor $\alpha_a$ & $0.0003$ \\
    Learning rate critic $\alpha_a$ & $0.001$ \\
    Discount factor $\gamma$ & 0.99 \\
    Beam tracking factor $\rho(k)$ & [3, 0, -1] \\
    Loss weighing $c_1$  & 0.5 \\
    Loss weighing $c_2$  & 0.01 \\
    Clipping parameter $\epsilon$ & 0.2 \\
    Optimizer & Adam \\
    \bottomrule
    \end{tabular}%
  \label{tab:_params}%
\end{table}%
\vspace{-5mm}
\section{Results}
\begin{figure}[ht]
    \centering
    \subfloat[ECDF for PER]{\includegraphics[width=0.35\textwidth]{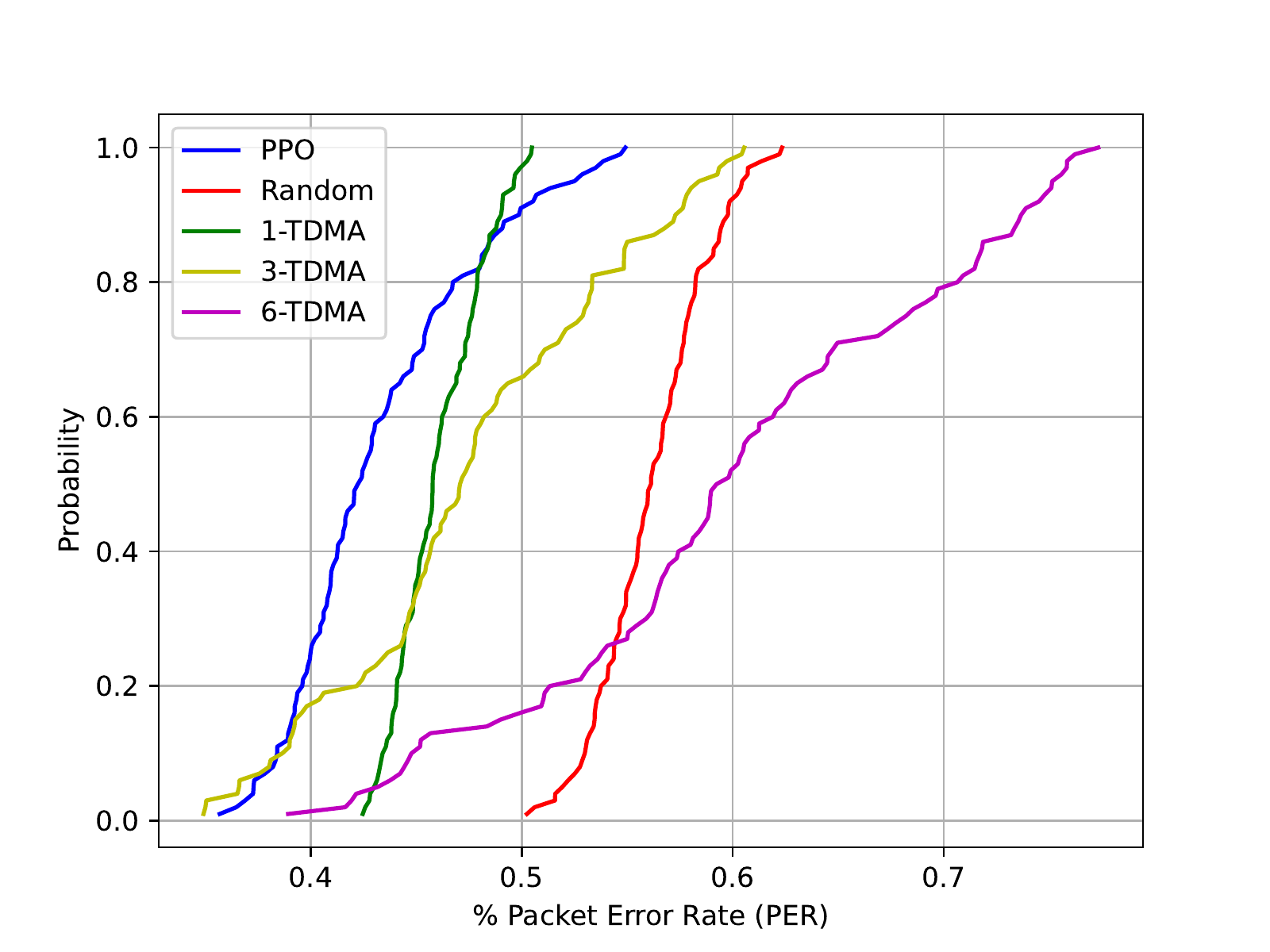}\label{fig:ecdf1}}
    \hfill
    \subfloat[ECDF for Throughput]{\includegraphics[width=0.35\textwidth]{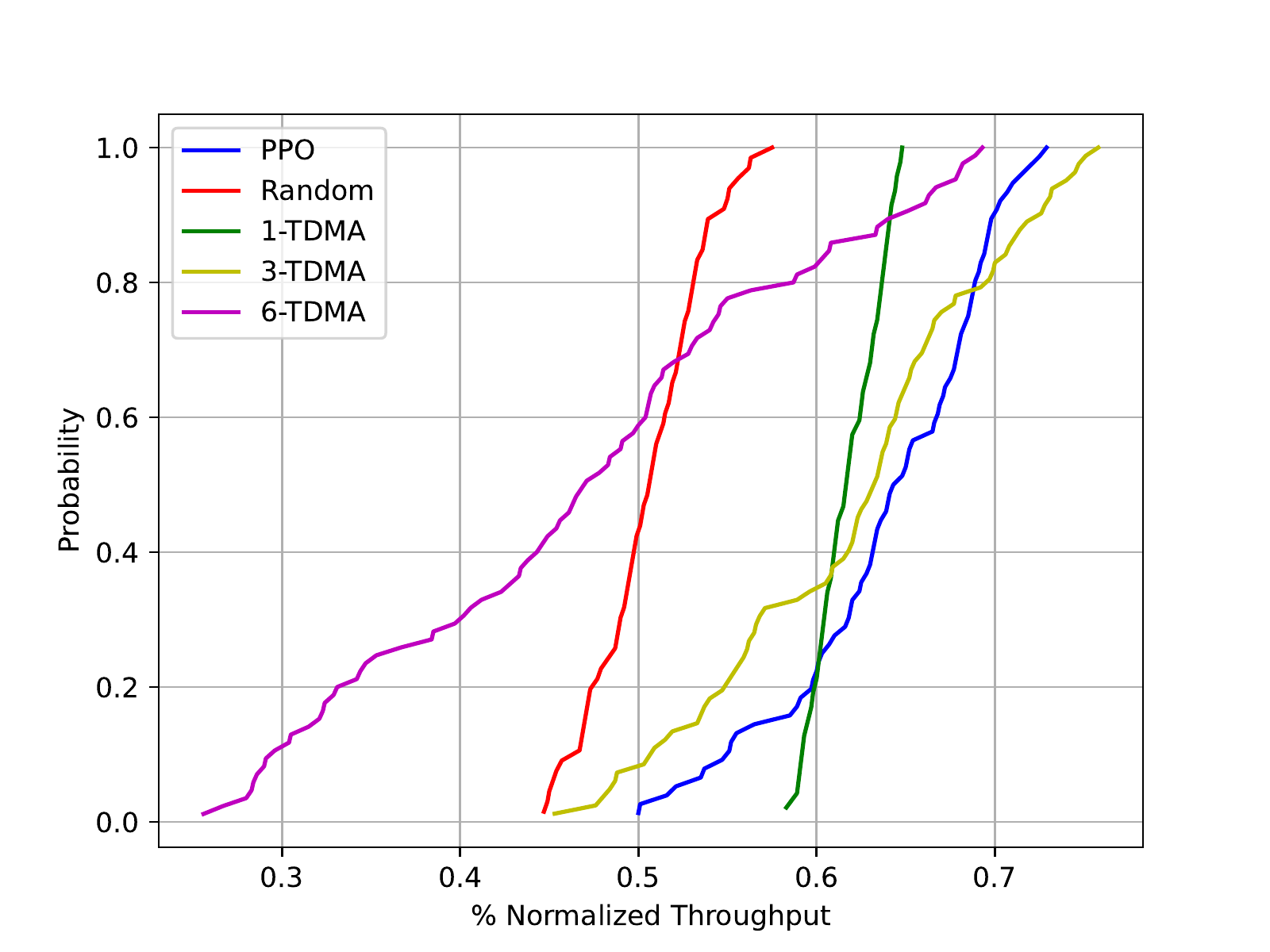}\label{fig:ecdf2}}
    \caption{ECDFs over $E_{test}$ episodes}
    \label{fig:ecdf}
\end{figure}

\begin{figure}[ht]
    \centering
    \includegraphics[width=0.6\columnwidth]{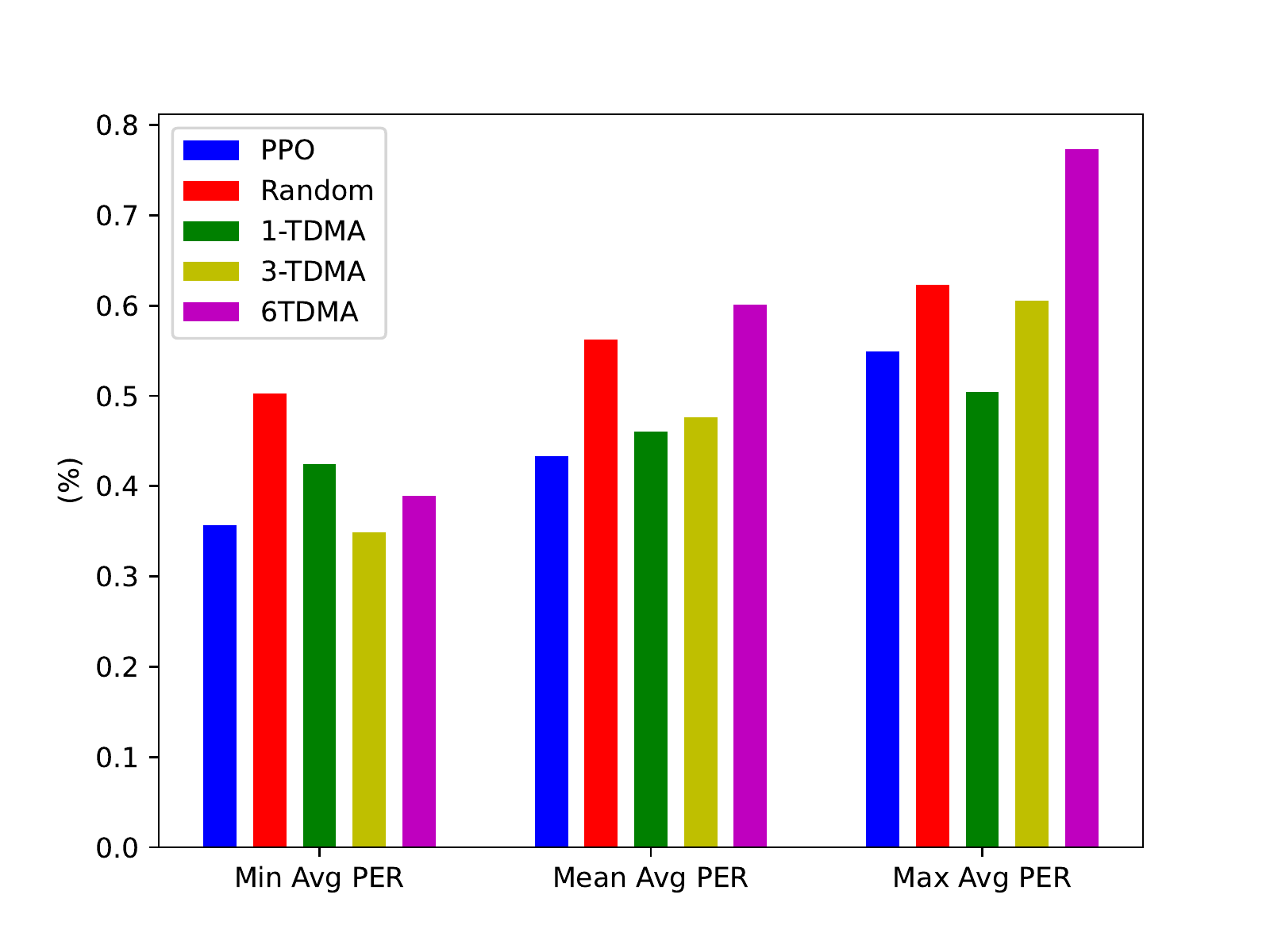}
    \caption{PER statistics over $E_{test}$ episodes}
    \label{fig:avgperstats}
\end{figure}

\begin{figure}[ht]
    \centering
    \includegraphics[width=0.6\columnwidth]{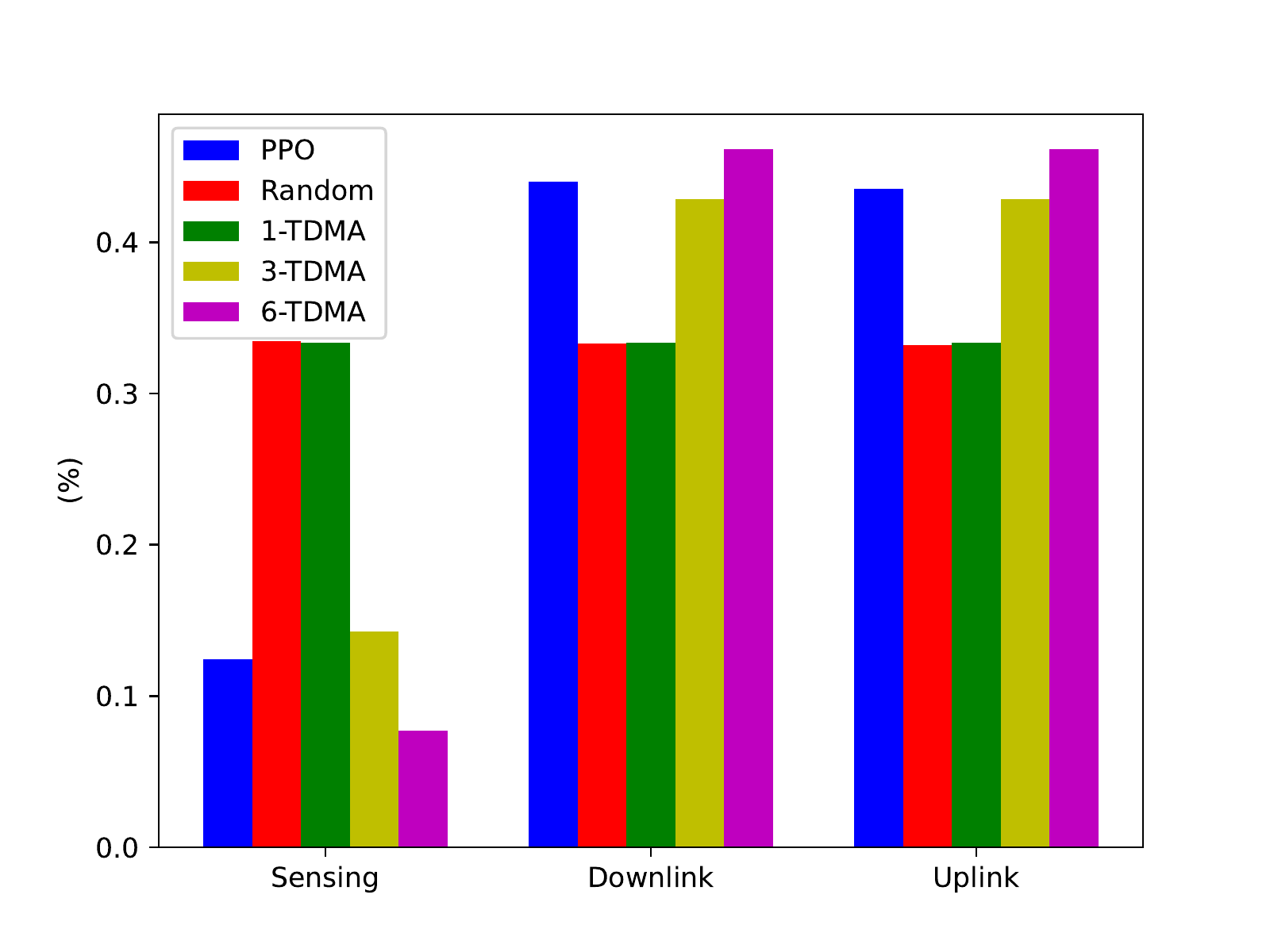}
    \caption{Slots distribution over $E_{test}$ episodes}
    \label{fig:slotsstats}
\end{figure}

\begin{figure}[ht]
    \centering
    \includegraphics[width=0.6\columnwidth]{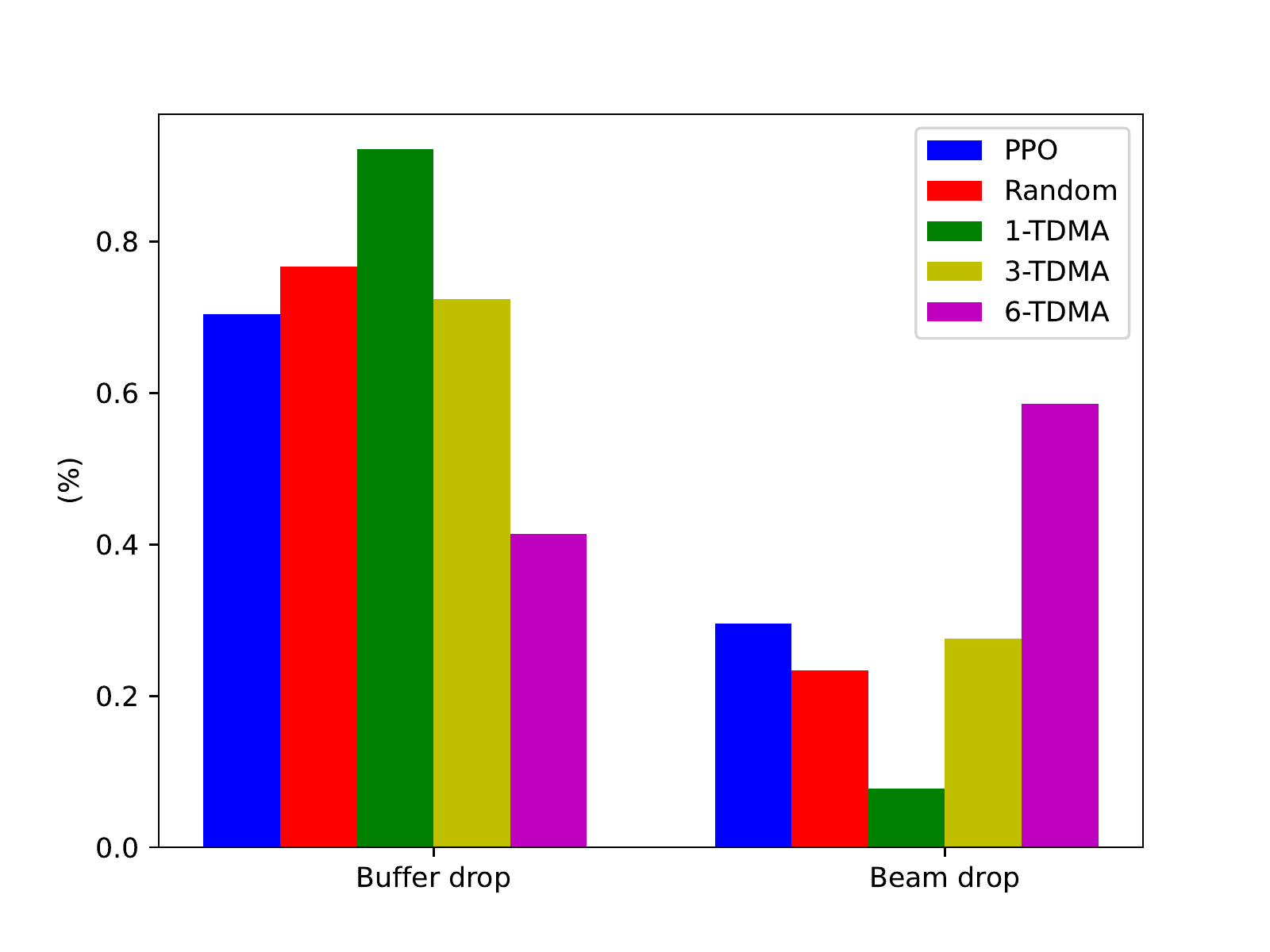}
    \caption{Drops distribution over $E_{test}$ episodes}
    \label{fig:drop_stats}
\end{figure}

We compare our proposed solution with the baselines in terms of average PER and normalized throughput across the evaluation episodes ($E_{\text{test}}$). Figure \ref{fig:ecdf} presents the empirical cumulative distribution function (ECDF) of the average PER and normalized throughput. Our method demonstrates effective policy learning compared to random allocation. It outperforms the baselines in terms of PER, even though 1-TDMA is the safest option on average, presenting a lower variance in the results. When reducing the number of sensing slots (3/6-TDMA), the variance of results increases significantly. Moreover, our method exhibits improved throughput performance. To provide a better understanding of our solution's effectiveness, Figure \ref{fig:avgperstats} illustrates the maximum, minimum, and average PER. The allocation policy of our method efficiently reduces the number of required sensing slots, as depicted in Figure \ref{fig:slotsstats}. Despite this reduction, our method achieves lower PER compared to the baselines, as observed in Figures \ref{fig:ecdf} and \ref{fig:avgperstats}. Furthermore, we analyze the efficiency in Figure \ref{fig:drop_stats}, which reveals that our dynamic allocation policy prioritizes transmitting packets to reduce the full buffer drop rate, rather than increasing sensing to minimize total PER but not as much as 6-TDMA does. In comparison to 1/3-TDMA, the dynamic allocation of transmissions provides flexibility to improve overall performance, as evidenced by the average PER results.
\vspace{-4mm}

\vspace{-3mm}
\section{Conclusion}
The proposed approach for joint resource allocation and beam alignment in mmWave wireless communication networks performs significantly better than baseline methods in terms of average packet error rate (PER). It provides an effective solution to the resource allocation problem and exhibits great generalization performance, regardless of system stochasticity. The approach is promising for integrated sensing and communications networks and we plan to explore a learned protocol design in future work.

\vspace{-2.3mm}
\bibliographystyle{IEEEtran}
\bibliography{reference.bib}

\end{document}

%% file: acronym.tex
\newacronym{AWGN}{AWGN}{additive white Gaussian noise}

\newacronym{bs}{BS}{base station}

\newacronym{DFT}{DFT}{discrete Fourier transform}
\newacronym{dl}{DL}{downlink}

\newacronym{emf}{EMF}{electromagnetic field}
\newacronym{emfe}{EMFE}{electromagnetic field exposure}

\newacronym{los}{LoS}{line-of-sight}

\newacronym{MIMO}{MIMO}{multiple-input, multiple-output}
\newacronym{MIMO-OFDM}{MIMO-OFDM}{multiple-input, multiple-output orthogonal frequency-division multiplexing}

\newacronym{MSE}{MSE}{mean square error}

\newacronym{NMSE}{NMSE}{normalized mean square error}

\newacronym{OFDM}{OFDM}{orthogonal frequency-division multiplexing}

\newacronym{ris}{RIS}{reflective intelligent surface}

\newacronym{SNR}{SNR}{signal-to-noise ratio}

\newacronym{ULA}{ULA}{uniform linear array}
\newacronym{upa}{UPA}{uniform planar array}
\newacronym{tdd}{TDD}{time division duplex}
\newacronym[plural = UEs]{ue}{UE}{user equipment}

\newacronym{5G}{5G}{5th-generation}
\newacronym{6G}{6G}{6th-generation}

%% file: letter.bbl
\begin{thebibliography}{1}
\providecommand{\url}[1]{#1}
\csname url@samestyle\endcsname
\providecommand{\newblock}{\relax}
\providecommand{\bibinfo}[2]{#2}
\providecommand{\BIBentrySTDinterwordspacing}{\spaceskip=0pt\relax}
\providecommand{\BIBentryALTinterwordstretchfactor}{4}
\providecommand{\BIBentryALTinterwordspacing}{\spaceskip=\fontdimen2\font plus
\BIBentryALTinterwordstretchfactor\fontdimen3\font minus
  \fontdimen4\font\relax}
\providecommand{\BIBforeignlanguage}[2]{{%
\expandafter\ifx\csname l@#1\endcsname\relax
\typeout{** WARNING: IEEEtran.bst: No hyphenation pattern has been}%
\typeout{** loaded for the language `#1'. Using the pattern for}%
\typeout{** the default language instead.}%
\else
\language=\csname l@#1\endcsname
\fi
#2}}
\providecommand{\BIBdecl}{\relax}
\BIBdecl

\bibitem{giordani2018tutorial}
M.~Giordani, M.~Polese, A.~Roy, D.~Castor, and M.~Zorzi, ``A tutorial on beam
  management for 3gpp nr at mmwave frequencies,'' \emph{IEEE Communications
  Surveys \& Tutorials}, vol.~21, no.~1, pp. 173--196, 2018.

\bibitem{susarla2022hierarchial}
P.~Susarla, Y.~Deng, M.~Juntti, and O.~S{\'\i}lven, ``Hierarchial-dqn
  position-aided beamforming for uplink mmwave cellular-connected uavs,'' in
  \emph{GLOBECOM 2022-2022 IEEE Global Communications Conference}.\hskip 1em
  plus 0.5em minus 0.4em\relax IEEE, 2022, pp. 1308--1313.

\bibitem{raj2022deep}
V.~Raj, N.~Nayak, and S.~Kalyani, ``Deep reinforcement learning based blind
  mmwave mimo beam alignment,'' \emph{IEEE Transactions on Wireless
  Communications}, vol.~21, no.~10, pp. 8772--8785, 2022.

\bibitem{ma2020machine}
W.~Ma, C.~Qi, and G.~Y. Li, ``Machine learning for beam alignment in millimeter
  wave massive mimo,'' \emph{IEEE Wireless Communications Letters}, vol.~9,
  no.~6, pp. 875--878, 2020.

\bibitem{wang2019mmwave}
Y.~Wang, A.~Klautau, M.~Ribero, A.~C. Soong, and R.~W. Heath, ``Mmwave
  vehicular beam selection with situational awareness using machine learning,''
  \emph{IEEE Access}, vol.~7, pp. 87\,479--87\,493, 2019.

\bibitem{liang2019deep}
L.~Liang, H.~Ye, G.~Yu, and G.~Y. Li, ``Deep-learning-based wireless resource
  allocation with application to vehicular networks,'' \emph{Proceedings of the
  IEEE}, vol. 108, no.~2, pp. 341--356, 2019.

\bibitem{sridhara2008spectrum}
K.~Sridhara, A.~Chandra, and P.~S. Tripathi, ``Spectrum challenges and
  solutions by cognitive radio: An overview,'' \emph{Wireless Personal
  Communications}, vol.~45, pp. 281--291, 2008.

\bibitem{schulman2017proximal}
J.~Schulman, F.~Wolski, P.~Dhariwal, A.~Radford, and O.~Klimov, ``Proximal
  policy optimization algorithms,'' \emph{arXiv preprint arXiv:1707.06347},
  2017.

\bibitem{rezaie2020location}
S.~Rezaie, C.~N. Manch{\'o}n, and E.~De~Carvalho, ``Location-and
  orientation-aided millimeter wave beam selection using deep learning,'' in
  \emph{ICC 2020-2020 IEEE International Conference on Communications
  (ICC)}.\hskip 1em plus 0.5em minus 0.4em\relax IEEE, 2020, pp. 1--6.

\end{thebibliography}
